\documentclass[twocolumn,showpacs,preprintnumbers,amsmath,amssymb,prl,tightenlines,final]{revtex4-1}

\usepackage{amsfonts,amssymb,amsmath,times}
\usepackage{hyperref,graphics} 

\usepackage{graphicx}
\usepackage{dcolumn}
\usepackage{bm}
\usepackage{epsfig}
\usepackage{longtable}

\input epsf

\begin{document}

\title{ Control of quantum interference in molecular junctions: Understanding the origin of Fano and anti- resonances } 
\date{\today} 
\author{Daijiro Nozaki}
\email{daijiro.nozaki@tu-dresden.de}
\author{H\^aldun Sevin\c{c}li}
\author{Stanislav M. Avdoshenko}
\author{Rafael Gutierrez}
\author{Gianaurelio Cuniberti}
\affiliation{Institute for Materials Science and Max Bergmann Center of Biomaterials, Dresden University of Technology, 01062 Dresden, Germany}


\begin{abstract}
\noindent
We investigate within a coarse-grained model the conditions leading to the appearance of Fano resonances or anti-resonances in the conductance spectrum of a generic molecular junction with a side group (T-junction). By introducing a simple graphical representation (parabolic diagram), we can easely visualize the relation between the different electronic parameters determining the regimes where Fano resonances or anti-resonances in the low-energy conductance spectrum can be expected.  The results obtained within the coarse-grained model are validated using density-functional based quantum transport calculations in realistic T-shaped molecular junctions.

\end{abstract}

\keywords{Quantum interference, Fano resonance, anti-resonance, electron transport}

\maketitle

{\textit{Introduction$-$}} Quantum interference (QI) effects in electron transport have  been broadly studied in the field of mesoscopic physics and quantum dots~\cite{RevModPhys,PhysRevB.67.195335}, and their appearance in  molecular junctions had been suggested in the early 1990s from  studies on  electron transfer~\cite{Ratner90,Ratner2}. In the context of charge transport in molecular scale electronic devices~\cite{rafa}, QI effects have drawn increasing attention over the past years due to their unique spectral features, manifested in the conductance spectra of the junctions, and which make them very attractive for tuning the low energy electrical response of molecular-scale systems~\cite{Emberly,Tada,walczak,Baranski,Sasada2005,PhysRevB.74.193306,075119,Gemma-JACS,PhysRevLett.103.266807,Stadler_JCP,Hansen,Markussen,Markussenpccp,doi:10.1021/nl201042m,doi:10.1021/nl901554s,doi:10.1021/nl201042m,Gema-beil}.
For instance, since QI effects introduce additional peaks or dips in the conductance spectrum,  it may be expected that they can considerably change  the on/off ratios and thermoelectric performance of molecular junctions~\cite{doi:10.1021/nl201042m,Stadler_JCP}.  
Moreover, since it is possible to infer from the analysis of transport line shapes the molecular electronic structure as well as the topological connectivity of the molecular junctions, QI related effects might be also exploited for sensor or interferometer applications by monitoring the changes of the transmission spectrum as function of e.g. a magnetic field.

 It is therefore of great theoretical and practical interest to provide simple rules controlling the emergence of different types of QI effects like anti-resonances or Fano line shapes. According to the lineshape, QI effects can manifest either as anti-resonances (negative, almost symmetric peaks) or as Fano resonances (asymmetric sharp peaks). As a general qualitative trend, it has been shown that T-shaped  or cyclic molecules  tend to exhibit QI effects in their transmission spectra. 

Although many studies  of QI effects have been conducted, 
the understanding of the origin of anti-resonances and Fano resonances as well as their classification  are  still under debate. 
For  further development of single molecular devices exploiting QI effects, the precise relationship between electronic structure and QI needs to be understood. In particular, the origins of and the constraints on Fano and anti-resonance line shapes should be clarified. Although this can be achieved using extensive first-principle based calculations, it is also desirable to provide simple rules relating electronic structure and QI signatures in molecular scale systems.  

Here, we first study the transmission properties of a well-known T-shaped toy molecule. Using it as a starting point, we describe the conditions under which the two types of QI line shapes (anti-resonances and Fano resonances) may appear. Hereby we introduce a simple graphical representation $-$called from now on parabolic diagram for the sake of simplicity, see also Fig.~\ref{fig2a}$-$ to predict  the type of QI effect as well as their line shape in an accurate way. 
In the second part of this study, we validate our minimal model approach by computing on a first-principle basis the linear conductance of realistic molecular junctions with different topologies and displaying QI related features, finding a very nice agreement with the parabolic diagram approach.   

\begin{figure}
\centering
\includegraphics[width =0.60\linewidth]{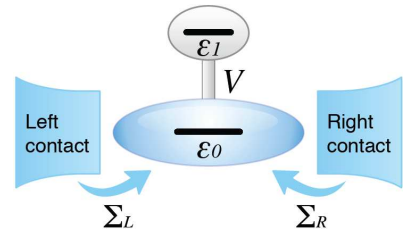}
\caption{Schematic model of a molecular junction with a side group considered in this study. A molecule consisting of two fragments is coupled between electrodes. The energies of two fragments  are $\varepsilon_0$ and $\varepsilon_1$. The transfer integral between them is set to $V$. $\Sigma_{\textrm{L/R}}$ represents the self energies due to coupling to the contacts.} 
\label{fig1}
\end{figure}

{\textit{Minimal model of a T-shaped molecule$-$}}We consider a generic molecular wire with a side group as schematically shown in Fig.~\ref{fig1}. The system is attached to left and right electronic baths only through the central wire, i.e. the side groups do not directly couple to the electrodes. To analyze QI effects in such a configuration, we can mainly focus on the low-energy (near the Fermi level $E_F$) features of the conductance spectrum. Hence, a simple coarse-grained can be formulated, which does not need to include the full electronic spectrum of the system. Consider the retarded Green's function $G^\textrm{r}$ of the molecule attached to the electrodes. It satisfies a spectral representation of the form~\cite{gdatta,Tada}: 
\begin{equation}
G^\textrm{r}(E)=\sum_{i=1}^{N}\phi_{i}\psi_{i}^{*}/(E+i\delta-\epsilon_{i}).
\label{eq0}
\end{equation}
Since the molecular system is coupled to a continuum of states (the electrodes) and is thus an open system, the eigenenergies $\epsilon_{i}$ are complex and the difference between left ($\phi_{i}$) and right ($\psi_{i}^{*}$) eigenvectors building a bi-orthonormal set should be taken into account. 
The transmission through the system around energies $E\sim E_F$, is mainly affected by few molecular orbitals (MOs) around $E$ since the contribution from deeper levels is expected to be small because of the large  denominators in eq.~(\ref{eq0}).
As a result, charge transport in the model system of Fig.~1 can be  simplified to a two-state model with energies $\varepsilon_0$ for the central site and $\varepsilon_1$ for the side group as long as we focus on the transmission  around these energies.  This two-site model is identical to the one modeled by Papadopoulos \textit{et al.}~\cite{PhysRevB.74.193306} and Stadler and Markussen \cite{Stadler_JCP}. The coupling between the two states is set to $V$. In real systems, $\varepsilon_0$ is expected to correspond to one of the frontier MOs, which is  delocalized over the main chain and is responsible for charge transport, while
 $\varepsilon_1$ corresponds to MOs of the side-functional group, which have the strongest overlap with the $\varepsilon_0$ states.

Within Landauer's theory~\cite{gdatta}, the transmission function can be computed via $T(E)=\textrm{Tr}[G^\textrm{r} \Gamma_\textrm{L}G^\textrm{a}\Gamma_\textrm{R}]$, where $\Gamma_{\textrm{L/R}}$ correspond to the spectral densities  of the electrodes. Since the coupling to the electrodes and to the side group can be represented by three self-energy terms, $\Sigma_{\textrm{L/R}}$ (due to the left/right reservoirs) and $\Sigma_{\textrm{S}}=V^2/(E-\varepsilon_1)$, the retarded Green's funcion of the molecular junction can be  written as: $[G^\textrm{r}]^{-1}(E)=(E+ i\delta)-\varepsilon_0-\Sigma_{\textrm{L}}-\Sigma_{\textrm{R}} -V^2/(E-\varepsilon_1)$.
If we neglect for simplicity the energy dependence of the electrode self-energies within the so called wide-band limit (WBL), the contribution of the reservoirs can be written as  $\Sigma_{\textrm{L/R}}=-i\gamma_{\textrm{L/R}}$. The broadening functions $\Gamma_{\textrm{L/R}}$ are then defined as $\Gamma_{\textrm{L/R}}\equiv i[\Sigma_{\textrm{L/R}}-\Sigma_{\textrm{L/R}}^{\dag}]=2\gamma_{\textrm{L/R}}$. Hereafter for simplicity, we assume $\gamma_{\textrm{L/R}}=\gamma$. 
The transmission function is then simply given by \cite{PhysRevB.74.193306,Stadler_JCP}:
\begin{equation}
T(E)=\frac{4\gamma^2}{(E-\varepsilon_0 -\frac{V^2}{E-\varepsilon_1})^2 + 4\gamma^2   }
\label{eq3}
\end{equation}

Although the same discussion is held in Ref. \cite{Stadler_JCP},
starting from this formula,  
we show below that,  
(i) an anti-resonance at $E = \varepsilon_1$ and two symmetric resonance peaks at  $\varepsilon_{\textrm{bond}}$ and $\varepsilon_{\textrm{anti}}$ appear when $|\varepsilon_1-\varepsilon_0|/|V|$ is small, (ii) the system presents asymmetric Fano line shapes when $|\varepsilon_1-\varepsilon_0|/|V|$ is large, using our parabolic model. The transition regime is however not universally defined, but depends on the specific values of the electronic parameters.  

From eq.~(\ref{eq3}), it is obvious that when the term $(E-\varepsilon_0 -V^2/(E-\varepsilon_1))^2$ is a minimum, $T(E)$ attains its maximum.  
This  is a Breit-Wigner-type resonant peak with width $\gamma$ \cite{Breit}. 
The positions of the peaks are  given by the solutions of the quadratic equation:
\begin{equation}
(E-\varepsilon_0)(E-\varepsilon_1)  -V^2 =0,
\label{eq4}
\end{equation} 
with solutions $\varepsilon_{\textrm{bond}},\varepsilon_{\textrm{anti}}$  
given by  $2\varepsilon_{\textrm{bond,anti}}=\varepsilon_0+\varepsilon_1\pm\sqrt{(\varepsilon_0-\varepsilon_1)^2+4V^2}$, 
where the transmission maxima take place~\cite{PhysRevB.74.193306,Stadler_JCP,075119,PhysRevB.67.195335}. Note that the two solutions are always real. 
At this point we introduce a graphical representation (parabolic diagram) by plotting the reduced variable $y=(E-\varepsilon_{0})(E-\varepsilon_{1})$ as a function of the energy $E$, as shown in  Fig.~\ref{fig2a}(a). Using this diagram, the peak and dip positions as well as their relation to the relevant  electronic parameters $\varepsilon_0, \varepsilon_1$, $V$, $\varepsilon_{\textrm{bond}},\varepsilon_{\textrm{anti}}$ can easily be visualized. 
\begin{figure}
\centering
\includegraphics[width =0.65\linewidth]{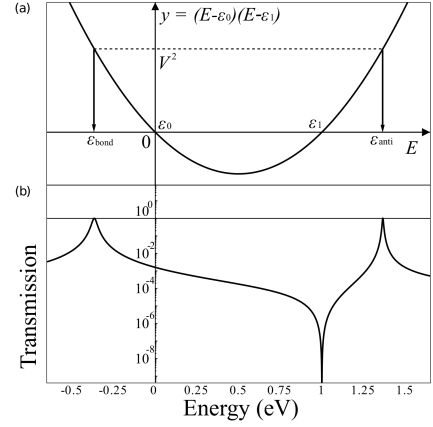}
\caption{ (a) Parabolic diagram showing the relation between on-site energies $\varepsilon_{0,1}$, coupling $V$ and resonant peaks $\varepsilon_{\textrm{bond,anti}}$. (b) The transmission function obtained from Eq.~(\ref{eq4}) with $\varepsilon_{0}=0.0$ eV, $\varepsilon_{1}=1.0$ eV, $V^2=0.5$ eV$^2$, and $ \gamma=0.01$ eV.   }
\label{fig2a}
\end{figure}
The system gives a dip at $E=\varepsilon_1$ with $T(E)=0$, since the denominator in eq.~(\ref{eq3}) diverges, $\lim_{E \to \varepsilon_1\pm0}\frac{V^2}{E-\varepsilon_1}=\pm\infty$ \cite{PhysRevB.74.193306,Hansen,Stadler_JCP}. 
Therefore, the energetic position of positive resonance peaks ($E=\varepsilon_{\textrm{bond}},\varepsilon_{\textrm{anti}}$) and negative peaks ($E=\varepsilon_1$) in $T(E)$  can be estimated from the  parabolic diagram in Fig.~\ref{fig2a}(a) without explicitly computing the transmission function from Eq.~(\ref{eq3}) or by using more sophisticated numerical methods. 
To illustrate this point further, if one would like to examine the QI effects of molecular wires having e.g. a conducting state at $\varepsilon_0=0.0$ eV and a localized state at $\varepsilon_1=1.0$ eV with coupling $V^2=0.5 $ eV$^2$, it is possible to estimate the position of the Breit-Wigner-type resonance and of an anti-resonance by drawing the diagram as in Fig.~\ref{fig2a}(a). Fig.~\ref{fig2a}(b) presents the transmission function calculated from eq.~(\ref{eq3}) using the same parameters. We can see that the position of the negative and positive peaks completely matches with the features of the parabolic diagram. 
More interestingly, it is also possible to estimate the coupling $V$ from the position of the positive and negative peaks in the transmission function using this parabolic diagram. If the peak ($\varepsilon_{\textrm{bond,anti}}$) and dip ($\varepsilon_1$) positions can be resolved experimentally, the orbital energy ($\varepsilon_0$) and the electronic coupling ($V$) are uniquely determined.

Next, we address the case of a Fano resonance. When one of the solutions of eq.~(\ref{eq4}) is close to $\varepsilon_1$, i.e. when $|\varepsilon_1-\varepsilon_0|/|V|$ is large enough, the negative and positive peaks come close together leading to an asymmetric line shape,  a Fano resonance. The condition for the anti-resonance is simply given by the $T(E)=0$ with large  energy spacing $|\varepsilon_{\textrm{anti}}-\varepsilon_1|$ or $|\varepsilon_{\textrm{bond}}-\varepsilon_1|$ with positive resonant peaks. This is satisfied by large $|V|$ with small $|\varepsilon_0-\varepsilon_1|$. In the special case of $\varepsilon_0=\varepsilon_1$, the anti-resonance shows a symmetric line shape at $E=\varepsilon_1$~\cite{Stadler_JCP,PhysRevB.67.195335}. 

\begin{figure}
\centering
\includegraphics[width =0.9\linewidth]{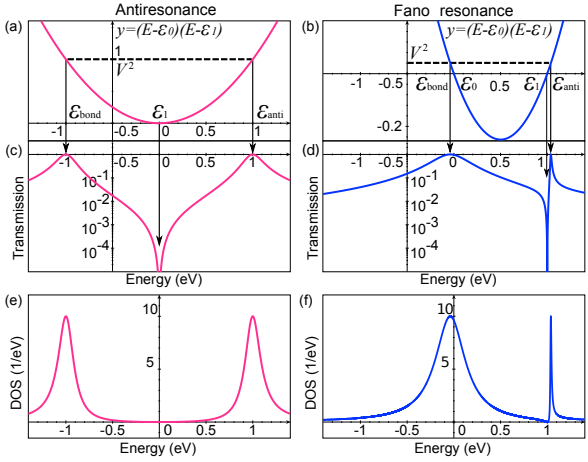}
\caption{The two types of QI through molecular junction in Fig.~\ref{fig1}. To produce the anti-resonance, parameters are set as $\varepsilon_{0,1}=0.0$ eV and $V=1.0$ eV. To produce the Fano resonance, parameters are set as $\varepsilon_{0}=0.0$ eV, $\varepsilon_{1}=1.0$ eV  and $V=0.2$ eV. }
\label{fig3a}
\end{figure}

{\textit{Minimal model of a T-shaped molecular junction: Dependence on the model parameters$-$}} Keeping the previously  discussed conditions for the two types of QI effects in mind, we will examine how  the transmission function behaves with different choices of $\varepsilon_0$, $\varepsilon_1$, and $V$ \textit{using the parabolic diagram}. Note that the investigation of the dependence on the site-energy and coupling constant is also done elsewhere \cite{PhysRevB.74.193306,Stadler_JCP}.  Fig.~\ref{fig3a} presents two types of QI effects of the system in Fig.~\ref{fig1}. As discussed before, small $|\varepsilon_0-\varepsilon_1|$ with large $|V|$ gives an anti-resonance (red in Fig.~\ref{fig3a}), while large $|\varepsilon_0-\varepsilon_1|$ with small $|V|$  gives a Fano resonance (blue in Fig.~\ref{fig3a}). The condition of weak coupling $|V|$ for the Fano resonance agrees with the general interpretation of the Fano effect where the localized state interferes with a continuum. A critical difference between these QI effects is that an anti-resonance does not require a localized state at the position of a negative peak, while a Fano resonance does require a localized state at the position of the asymmetric peak.

\begin{figure}
\centering
\includegraphics[width =1\linewidth]{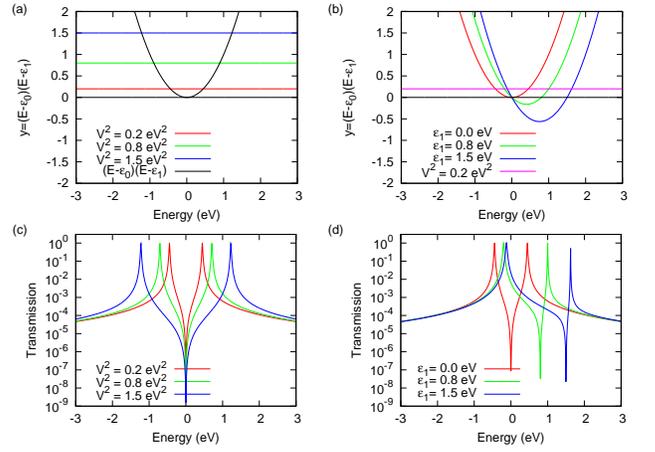}
\caption{ (c) $|V|$ dependence of transmission function of the system in Fig.~\ref{fig1} and (a) the corresponding parabolic diagram. (d) On-site energy dependence of the transmission function of the system in Fig.~\ref{fig1} and  (b) the corresponding parabolic diagram. In (a) and (c), the on-site energies $\varepsilon_{0,1}$ are set as $\varepsilon_{0,1}=0.0$ eV. In (b) and (d), the on-site energy for $\varepsilon_{0}$ is set as $\varepsilon_{0}=0.0$ eV. The coupling strength $|V|$ is fixed to $V^2=0.2$ eV$^2$. The transmission spectra gradually shifts from  anti-resonance having symmetric negative peaks  to Fano resonance having asymmetric peaks as  $|\varepsilon_0-\varepsilon_1|$ is increased. } 
\label{fig4}
\end{figure}

We also investigated the coupling $|V|$ dependence and on-site energy $|\varepsilon_1|$ dependence  of the QI effects. Fig.~\ref{fig4}(c) presents the anti-resonance with different coupling strengths. Fig.~\ref{fig4}(a) is corresponding parabolic diagram.  In the weak coupling limit, the transmission merges down to a
Breit-Wigner resonance. This is obvious from Eq.~(\ref{eq4}) and also from the parabolic diagram since the resonant peaks come closer to each other with decreasing $V$. 
Fig.~\ref{fig4}(d) shows the on-site energy $|\varepsilon_1|$ dependence of the transmission. Fig.~\ref{fig4}(b) is corresponding parabolic diagram for the on-site energy dependence. We can see the transition from anti-resonance to Fano resonance with larger $|\varepsilon_1-\varepsilon_0|$. 
This is also interpreted from the parabolic diagram in Fig.~\ref{fig4}(b) since the gradient of the parabola at $E=\varepsilon_1$  becomes steeper for increasing $|\varepsilon_1-\varepsilon_0|$, which makes $\varepsilon_1$ and $\varepsilon_{\textrm{anti}}$ closer to each other.   From these analysis, we conclude that weak coupling $|V|$ and weak broadening $\gamma$ with large $|\varepsilon_0-\varepsilon_1|$ are the optimal conditions for  Fano resonance. 

\begin{figure}[ht!]
\centering
\includegraphics[width =0.9\linewidth]{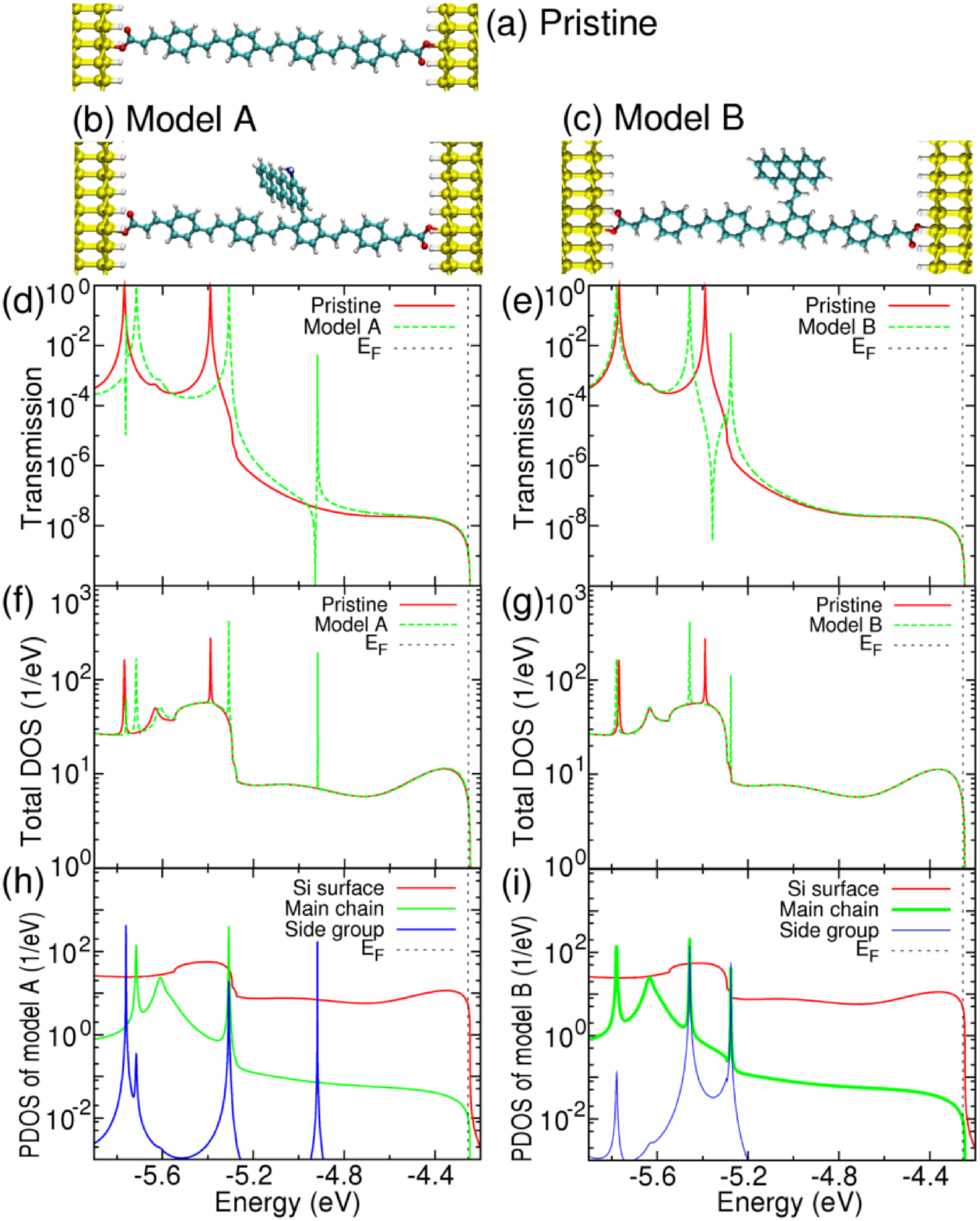}
\caption{ Demonstration of Fano resonance (left panels) and antireonance (right panels). } 
\label{fig5}
\end{figure}

{\textit{Validation of the parabolic diagram: first-principle transport calculations$-$}}
In order to demonstrate the relevance of the parabolic diagram  for realistic systems, we have modeled two  molecular wires with side groups as shown in Fig.~\ref{fig5}, which display  Fano resonances and anti-resonances. As electrodes, we chose semi-infinite Si slabs with a (111) termination and surface dangling bonds saturated with hydrogens. We used  the gDFTB program~\cite{gdftb}  for the transmission calculations, which is based on a  density-functional tight binding method~\cite{dftb}. All transmission calculations were performed after structure relaxation in the same way as in Ref.~\cite{switch} 

At first, in order to generate a Fano resonance, we modeled a molecular wire (model A in Fig.~\ref{fig5}(b)) by weakly coupling  a side-group to the main chain such that the energy level of the eigenstate responsible for the conduction ($\varepsilon_{\textrm{anti}}$)  is separated enough to that of the localized fragment MO at the side group ($\varepsilon_1$) . 
To have weak electronic coupling between them, the side group is vertically grafted to the main chain. We used carboxylic acids for the linkers connecting between the molecule and the Si contacts.    Fig.~\ref{fig5} summarizes the transmission and total (DOS) and projected (PDOS) density of states  plots of the modeled systems.   In the absence of the side group, the valley with low transmission spreads from the HOMO to the Fermi energy, while in the presence of the side group a Fano resonance appears around $E=-4.9$ eV  due to the interference between the localized state from the side group and the conduction state.  We can clearly see the localized state around $E=-4.9$ eV originating from the side group in PDOS plot in Fig.~\ref{fig5}(d). The coupling strength $V$ and  the energy difference $|\varepsilon_0-\varepsilon_1|$ are estimated to be 0.062 eV and 0.370 eV from the transmission line shape in Fig.\ref{fig5}(d). As mentioned in the analysis of the coarse-grained model, these values with large $|\varepsilon_1-\varepsilon_0|/|V| = 5.968$ satisfy the condition for Fano resonance.

Next, in order to have an anti-resonance, we modeled a molecular wire (model B Fig.~\ref{fig5}(c)) by strongly coupling a side-group to the main chain. 
For strong electronic coupling between them, the side group is linked to the main chain using an $sp_2$ linker. 
We can see that the resonant peak of the pristine system at $-5.4$ eV in Fig.~\ref{fig5}(e) splits into bonding and anti bonding peaks after the coupling to the side group because of the strong electronic interaction between the side group and the main chain. The transmission of model B gives an anti-resonance peak  between the bonding and anti bonding peaks as expected. 
From the PDOS plot in Fig.~\ref{fig5}(i), it is clear that the states of the side group are not localized anymore (thus not satisfying the condition for Fano resonance) but delocalized on the main chain because of  the strong electronic interaction between the side group and the main chain.
The coupling strength $V$ and  the energy difference $|\varepsilon_0-\varepsilon_1|$ are estimated to be 0.090 eV and 0.023 eV from the transmission function in Fig.~\ref{fig5}(e). As mentioned in the analysis of the coarse-grained model, these values with large $V$ compared to $|\varepsilon_0-\varepsilon_1|$ satisfy the condition for anti-resonance ( $|\varepsilon_1-\varepsilon_0|/|V| = 0.256$ ).

{\textit{Conclusion$-$}} We have derived conditions for Breit-Wigner type normal resonant peaks as well as for quantum intereference related states (Fano resonances and anti-resonances) in T-shaped molecular junctions using a coarse-grained model and clarified the origins of the two QI-based effects using a simple graphical representation (parabolic diagram). 
With its help, we can easely explain the two QI related effects in realistic molecular junctions, thus validating our minimal formulation. Our results are expected to  provide a very helpful guideline for building functional molecular devices which exploit QI effects. 
Usually, in planar organic $\pi$-systems,  the electronic coupling $V$ between the main molecular chain and the side groups is strong. This situation is closer to the condition for anti-resonance than for a Fano resonance and, hence, this is the reason why many organic systems display anti-resonances while Fano resonances are rarely seen. 

TThis work is funded by the European Union (ERDF) and the Free State of Saxony via TP A2 ("MolFunc") of the cluster of excellence "European Center for Emerging Materials and Processes Dresden" (ECEMP), by the European Union (ERDF) and the Free State of Saxony via the ESF project 080942409 InnovaSens, and by World Class University program funded by the Ministry of Education, Science and Technology through the National Research Foundation of Korea (R31-10100).
 \\

\bibliographystyle{apsrev4-1}

\end{document}